\documentclass[12pt,a4paper]{article}
\usepackage{jheppub}

\usepackage{amsmath,amssymb}
\usepackage{hyperref}

\newcommand{\trans}{{\scriptscriptstyle\mathsf{T}}}
\newcommand{\grp}[1]{\ensuremath\mathrm{#1}}

\DeclareMathOperator{\tr}{tr}
\DeclareMathOperator{\sdet}{sdet}

\title{Twistor Parametrization of Locally BPS Super-Wilson Loops}
\author{C.~Vergu}
\affiliation{Department of Mathematics, King's College London \\
The Strand, WC2R 2LS, London, UK}
\emailAdd{c.vergu@gmail.com}

\abstract{We consider the kinematics of the locally BPS super-Wilson loop in \(\mathcal{N}=4\) super-Yang-Mills with scalar coupling from a twistorial point of view.  We find that the kinematics can be described either as supersymmetrized pure spinors or as a point in \(G_{2\vert 2}(4\vert 4) \times G_{2\vert 2}(4\vert 4)\).  In this description of the kinematics the scalar--scalar correlation function appearing in the one-loop evaluation of the super-Wilson loop can be neatly written as a sum of four super-determinants.}

\begin{document}

\maketitle

\section{Introduction}
\label{sec:introduction}

The dual superconformal symmetry is one of the more surprising and unexpected results arising from studies of scattering amplitudes in \(\mathcal{N} = 4\) theory.  It started as an observation (see ref.~\cite{Drummond:2006rz}) of an inversion symmetry property of integrals appearing the perturbative expansion.  From there it grew into a full superconformal symmetry, as shown in ref.~\cite{Drummond:2008vq} and finally into an infinite-dimensional Yangian in ref.~\cite{Drummond:2009fd}.

The Yangian symmetry is at the core of the duality between polygonal super-Wilson loops and scattering amplitudes; in particular it contains both the conformal group acting on scattering amplitudes and the dual conformal group acting on super-Wilson loops.

Initially the duality was between polygonal Wilson loops and MHV scattering amplitudes (see refs.~\cite{Drummond2008c, Brandhuber2008a, Drummond2008b, Bern2008a, Drummond:2008aq}).  Later it was extended beyond MHV for scattering amplitudes and to polygonal \emph{super}-Wilson loops (see refs.~\cite{Mason2010, Caron-Huot2011}). However, it has proven challenging to actually use this big symmetry to constrain the functional form of scattering amplitudes or of super-Wilson loops, mainly because of IR/UV divergences (see however refs.~\cite{Drummond2010b, Caron_Huot_2011, Bullimore:2011kg, Caron-Huot2012, Beisert2012a, Beisert2012} for progress in this area).

For the super-Wilson loops the divergences arise because of the cusps in the contour.  Ref.~\cite{M_ller_2013} asked the question if it was possible to avoid the cusps (and the divergences) while still preserving the Yangian symmetry.  They found that this is possible but, besides the coupling to the gauge field \(A\) they had to add a coupling the scalars \(\phi\) as well, in a way which would correspond to the dimensional reduction of a light-like Wilson loop in ten-dimensional \(\mathcal{N} = 1\) super-Yang-Mills theory.  For a \(\grp{SU}(N)\) gauge theory, this super-Wilson loop reads
\begin{equation}
  \mathcal{W} = \frac 1 N \left\langle \tr P \exp\left(\int_C d t (p \cdot A + \dot{\theta} \cdot A_\theta + \dot{\bar{\theta}} \cdot A_{\bar{\theta}} + q \cdot \Phi)\right)\right\rangle,
\end{equation}
where \(p = \dot{x} + \tfrac 1 2 \dot{\theta} \bar{\theta} - \tfrac 1 2 \theta \dot{\bar{\theta}}\), \(q\) is a six-dimensional vector such that \(p^2 - q^2 = 0\), \(A\) is the Grassmann even gauge superfield, \(A_\theta\) and \(A_{\bar{\theta}}\) are the Grassmann-odd components of the gauge superfield, and \(\Phi\) is the scalar superfield.

Such Wilson loops were studied soon after the formulation of the AdS/CFT correspondence, in refs.~\cite{Rey2001, PhysRevLett.80.4859}. Their supersymmetrization was studied in ref.~\cite{Ooguri2000}.

The authors of ref.~\cite{M_ller_2013} found that the Yangian symmetry is present to one loop order for the first few orders in the Grassmann expansion.  Later, ref.~\cite{Beisert:2015jxa} studied the superconformal symmetry, \(\kappa\) symmetry and finiteness of these operators while ref.~\cite{Beisert:2015uda} studied the Yangian symmetry to one loop order but to all orders in the Grassmann expansion.

In ref.~\cite{Beisert:2015jxa} the scalar--scalar two-point function was computed with the result\footnote{Here and in the following we adopt a matrix notation, which allows us to omit all the indices.  The index contraction is just matrix multiplication or matrix trace.  In this notation \(x\) is a \(2 \times 2\) matrix, \(\theta\) is a \(2 \times 4\) matrix, \(\bar{\theta}\) is a \(4 \times 2\) matrix.  The conventions are similar to the ones in refs.~\cite{Beisert:2015uda, Beisert:2015jxa}.}
\begin{equation}
  \langle \Phi^{a b}(1) \bar{\Phi}_{c d}(2)\rangle \propto \frac {(1 + \bar{\theta}_{21} x_{21}^{+-,-1} \theta_{21})_{[c}{}^a (1 + \bar{\theta}_{21} x_{21}^{+-,-1} \theta_{21})_{d]}{}^b}{\det x_{21}^{+-}},
\end{equation}
where point \(i\) has coordinates \((x_i, \theta_i, \bar{\theta}_i)\) and \(\theta_{12} = \theta_1 - \theta_2\), \(x_{12}^{+-} = x_1^+ - x_2^- - \theta_1 \bar{\theta}_2\).

This result, while not difficult to obtain, does not make manifest the superconformal symmetry.  Ideally one would like to find a set of variables with simple superconformal transformations, which can be used as building blocks for superconformal and Yangian invariants.\footnote{In fact, we actually only need invariance up to total derivatives.}  This is what we set out to do in this paper.

We describe the kinematics by two \((2\vert 2) \times (2 \vert 4\vert 2)\) super-matrices \(\mathcal{V}\) and \(\tilde{\mathcal{V}}\) and two \((2\vert 4\vert 2) \times (2\vert 2)\) super-matrices \(\mathcal{U}\) and \(\tilde{\mathcal{U}}\), such that
\begin{equation}
  \mathcal{V} \tilde{\mathcal{U}} = 0, \qquad
  \tilde{\mathcal{V}} \mathcal{U} = 0, \qquad
  \sdet(\tilde{\mathcal{V}} \; \cdot) = \sdet(\mathcal{U}; \cdot)^{-1}, \qquad
  \sdet(\mathcal{V} \; \cdot) = \sdet(\tilde{\mathcal{U}}; \cdot)^{-1},
\end{equation}
where the \(\cdot\) can be replaced by any \((2\vert 4\vert 2) \times (2\vert 2)\) super-matrix such that all the super-determinants make sense.  The \(\mathcal{V}\) and \(\tilde{\mathcal{V}}\) transform by a right action of the superconformal group and are defined up to a left action by \((2\vert 2) \times (2\vert 2)\) super-matrices.  The same holds for \(\mathcal{U}\) and \(\tilde{\mathcal{U}}\) but with left and right interchanged.

Using these variables, the scalar--scalar correlation function can be written as
\begin{equation}
  \langle q_1 \cdot \Phi(1)\; \bar{q}_2 \cdot \bar{\Phi}(2)\rangle \propto \sdet (\mathcal{V}_1 \mathcal{U}_2) + \sdet (\mathcal{V}_1 \tilde{\mathcal{U}}_2) + \sdet (\tilde{\mathcal{V}}_1 \mathcal{U}_2) + \sdet (\tilde{\mathcal{V}}_1 \tilde{\mathcal{U}}_2).
\end{equation}

We hope that this parametrization, besides making the superconformal symmetry manifest, will be useful in finding the Yangian invariants as well, along the lines of refs.~\cite{ArkaniHamed:2009dn, ArkaniHamed:2009vw, Mason:2009qx}.

\section{Kinematics}
\label{sec:kinematics}

For a \(2 n \times 2 n\) anti-symmetric complex matrix \(A\), there is a decomposition \(A = U \Sigma U^\trans\), where \(U\) is a unitary matrix and \(\Sigma\) is a block-diagonal anti-symmetric matrix with \(2 \times 2\) blocks
\begin{equation}
  \Sigma =
  \begin{pmatrix}
    0 & c_1 & 0 & 0 & \hdotsfor{2}\\
    -c_1 & 0 & 0 & 0 & \hdotsfor{2}\\
    0 & 0 & 0 & c_2 & \hdotsfor{2}\\
    0 & 0 & -c_2 & 0 & \hdotsfor{2}\\
    \vdots & \vdots & \vdots & \vdots & \ddots\\
    \cdots & 0 & 0 & 0 & 0 & c_n\\
    \cdots & 0 & 0 & 0 & -c_n & 0
  \end{pmatrix}.
\end{equation}
Moreover, we can choose the \(c_i\) to be real and positive.

A similar decomposition exists for real anti-symmetric matrices, for which the unitary matrices in the decomposition above are replaced by orthogonal matrices.

In the following we consider \(q_{ab}\), which is a \(4 \times 4\) anti-symmetric complex matrix (therefore we set \(n = 2\) in the general discussion above).  We denote by \(\overline{q}\) the complex conjugate matrix.  Its decomposition reads \(\overline{q} = U^* \Sigma U^\dagger\).

The matrix \(q\) can be identified with a six-dimensional vector.  We denote the norm of this vector by \(q^2\).  The matrices \(q\) and \(\overline{q}\) are related by a constraint \(q \overline{q} = -q^2 \mathbf{1}_4\).  Using the decomposition above, we find that this implies \(q^2 = c_1^2 = c_2^2\) and, since \(c_1\) and \(c_2\) are real and positive, \(c_1 = c_2\).  We denote the common value by \(c\).

The final constraint we impose is the duality condition \(q_{a b} = \tfrac 1 2 \epsilon_{a b c d} \overline{q}^{c d}\).  It implies that \(\det U = 1\), therefore we have \(U \in \grp{SU}(4)\) while initially we had \(U \in \grp{U}(4)\).

In conclusion, \(q = U \Sigma U^\trans\) with \(U \in \grp{SU}(4)\) and
\begin{equation}
  \Sigma =
  \begin{pmatrix}
    0 & c & 0 & 0\\
    -c & 0 & 0 & 0\\
    0 & 0 & 0 & c\\
    0 & 0 & -c & 0
  \end{pmatrix},
\end{equation}
and \(c > 0\).

Given \(q\), there are several matrices \(U\) which yield the same \(q\).  For example, if \(P \Sigma P^\trans = \Sigma\), \(U\) and \(U' = U P\) give the same \(q\).  Such matrices \(P\) form a group, which is \(\grp{Sp}(4, \mathbb{C})\).  If we also want to preserve the unitarity conditions we should take \(P \in \grp{Sp}(4, \mathbb{C}) \cap \grp{U}(4)\).  This is a real form\footnote{The group \(\grp{Sp}(2)\) can be also defined as the subgroup of \(\grp{GL}(2, \mathbb{H})\) which preserves the hermitian form \(\langle x, y\rangle = \sum_{i=1}^2 \bar{x}_i y_i\), where \(x_i, y_i \in \mathbb{H}\) and \(\overline{x}_i\) is the quaternionic conjugate of \(x_i\).} of \(\grp{Sp}(4, \mathbb{C})\), called \(\grp{Sp}(2)\) (or sometimes \(\grp{USp}(4)\)).

This means that \(q\) is parametrized by the length \(q^2 = c^2\) and by the coset \(\grp{SU}(4)/\grp{Sp}(2)\).  Using the isomorphisms \(\grp{SU}(4) \simeq \grp{Spin}(6)\), \(\grp{Sp}(2) \simeq \grp{Spin}(5)\), \(\grp{SO}(5) \simeq \grp{Sp}(2)/\mathbb{Z}_2\), \(\grp{SO}(6) \simeq \grp{SU}(4)/\mathbb{Z}_2\), the coset is seen to be a five-dimensional sphere
\begin{equation}
  \grp{SU}(4)/\grp{Sp}(2) \simeq \grp{Spin}(6)/\grp{Spin}(5) \simeq \grp{SO}(6)/\grp{SO}(5) \simeq \mathbb{S}^5.
\end{equation}

A conformal transformation \(g\) can be seen as an element of \(\grp{SU}(2, 2)\).  We write
\begin{equation}
  g =
  \begin{pmatrix}
    A & B\\ C & D
  \end{pmatrix},
\end{equation}
where \(A\), \(B\), \(C\) and \(D\) are \(2 \times 2\) matrices.  The block decomposition of the inverse is
\begin{equation}
  g^{-1} =
  \begin{pmatrix}
    \tilde{A} & \tilde{B}\\ \tilde{C} & \tilde{D}
  \end{pmatrix}.
\end{equation}

The condition that \(g \in \grp{SU}(2, 2)\) is
\begin{equation}
  \begin{pmatrix}
    A & B\\ C & D
  \end{pmatrix}
  \begin{pmatrix}
    0 & 1\\-1 & 0
  \end{pmatrix}
  \begin{pmatrix}
    A^\dagger & C^\dagger\\ B^\dagger & D^\dagger
  \end{pmatrix} =
  \begin{pmatrix}
    0 & 1\\ -1 & 0
  \end{pmatrix},
\end{equation}
which implies that
\begin{equation}
  \begin{pmatrix}
    \tilde{A} & \tilde{B}\\ \tilde{C} & \tilde{D}
  \end{pmatrix} =
  \begin{pmatrix}
    D^\dagger & -B^\dagger\\ -C^\dagger & A^\dagger
  \end{pmatrix}.
\end{equation}

The complexified and compactified Minkowski space is the \(G_2(4)\) Grassmannian of two-planes through the origin of \(\mathbb{C}^4\).  A point on the Grassmannian can be described as a \(4 \times 2\) matrix of rank two modulo the right action of \(\grp{GL}(2)\).  In some appropriate coordinate patch, a representative of this coset can be written as \(\left(\begin{smallmatrix} x\\1\end{smallmatrix}\right)\). The \(2 \times 2\) matrix \(x\) encodes the space-time coordinates. In \((1,3)\) Lorentzian signature \(x\) is hermitian.

A conformal transformation \(g\) described by a \(4 \times 4\) matrix above, acts on \(\mathbb{C}^4\), the space on which the Grassmannian is defined, by left multiplication.  If we perform such a transformation followed by a right multiplication by \(\grp{GL}(2)\) to go back to the canonical form, we find the transformation law
\begin{equation}
  \begin{pmatrix}
    x\\1
  \end{pmatrix} \to
  \begin{pmatrix}
    A & B\\C & D
  \end{pmatrix}
  \begin{pmatrix}
    x\\1
  \end{pmatrix} \sim
  \begin{pmatrix}
    (A x + B)(C x + D)^{-1}\\1
  \end{pmatrix} =
  \begin{pmatrix}
    x'\\1
  \end{pmatrix}.
\end{equation}
Therefore, \(x \to x' = (A x + B)(C x + D)^{-1}\).

Let us also consider a dual \(G_2(4)\) Grassmannian, which is the set of two-planes through the dual \(\mathbb{C}^4\) to the one previously considered.  A point in this Grassmannian can be represented by a \(2 \times 4\) matrix, modulo a left action by \(\grp{GL}(2)\).  The conformal group acts on the right, by multiplication by the inverse. A representative of this coset can be written as \(\left(\begin{smallmatrix} 1& -x\end{smallmatrix}\right)\).  Under conformal transformations we have
\begin{equation}
  \begin{pmatrix}
    1 & -x
  \end{pmatrix} \to
  \begin{pmatrix}
    1 & -x
  \end{pmatrix}
  \begin{pmatrix}
    \tilde{A} & \tilde{B}\\
    \tilde{C} & \tilde{D}
  \end{pmatrix} \sim
  \begin{pmatrix}
    1 & -(\tilde{A} - x \tilde{C})^{-1}(-\tilde{B} + x \tilde{D})
  \end{pmatrix} =
  \begin{pmatrix}
    1 & -x'
  \end{pmatrix}
\end{equation}
We therefore have two ways to write the transformations of \(x\): \(x \to x' = (A x + B)(C x + D)^{-1}\) and \(x \to x' = (-x \tilde{C} + \tilde{A})^{-1} (x \tilde{D} - \tilde{B})\).  These two ways are related by hermitian conjugation.

For a \(2 \times 2\) matrix \(A\) we define \(\overline{A} = (\det A) A^{-1} = -\epsilon A^\trans \epsilon\).  This operation has the following properties:
\(\overline{A B} = \overline{B}\; \overline{A}\),
\(\overline{(A^{-1})} = (\overline{A})^{-1}\),
\(\det A = \det \overline{A}\) and
\(\overline{\overline{A}} = A\).

The \(2 \times 2\) matrix \(p\) is hermitian since \(x\) is hermitian. Therefore, \(p\) can be diagonalized by a unitary matrix \(v\):
\begin{equation}
  p = v
  \begin{pmatrix}
    c_1 & 0\\ 0 & c_2
  \end{pmatrix} v^\dagger.
\end{equation}

The unitary matrix \(v\) can be restricted to be in \(\grp{SU}(2)\) (in fact, even in \(\grp{PSU}(2)\)) and we will do so from now on. Since \(\det v = 1\) and \(v\) is \(2 \times 2\), we have \(v^{-1} = v^\dagger = -\epsilon v^\trans \epsilon\) and also
\begin{equation}
  p \epsilon = v
  \begin{pmatrix}
    0 & c_1\\ -c_2 & 0
  \end{pmatrix} v^\trans, \qquad
  \overline{p} \epsilon = v
  \begin{pmatrix}
    0 & c_2\\ -c_1 & 0
  \end{pmatrix} v^\trans, \qquad
  \overline{p} = v
  \begin{pmatrix}
    c_2 & 0\\ 0 & c_1
  \end{pmatrix} v^\dagger,
\end{equation}
which shows that \((p \epsilon)^\trans = -(\overline{p} \epsilon)\). If we denote by \(p^2\) the norm of the momentum, then \(p^2 = c_1 c_2 = \det p = \det \overline{p}\) and \(p^0 = \tfrac 1 2 \tr p = \tfrac 1 2 \tr \overline{p}\).

Given \(p\), \(v\) is not uniquely determined.  In fact, \(p\) can be parametrized by \(c_1\), \(c_2\) and a coset \(\mathbb{S}^2 \simeq \grp{SU}(2)/\grp{U}(1)\).  In other words, the momentum is parametrized by its zero component, its norm and the direction of its space-like projection.

\section{Pure Spinor Parametrization}
\label{sec:pure-spinor-param}

Let us now solve the following equation in \(\lambda\)
\begin{equation}
  \lambda \bar{p} \epsilon - q \lambda^* = 0.
\end{equation}
Using the decomposition of \(\bar{p}\) and \(q\) worked out above we can reduce this equation to
\begin{equation}
  \xi
  \begin{pmatrix}
    0 & -c_2\\
    c_1 & 0
  \end{pmatrix} + c
  \begin{pmatrix}
    \epsilon & 0\\0 & \epsilon
  \end{pmatrix} \xi^* = 0.
\end{equation}
This equation needs to be solved in the \(4 \times 2\) matrix \(\xi\). Given a solution of this equation we can find \(\lambda = U \xi v^\dagger\).

When \(c_1 c_2 = c^2\), which just corresponds to a 10D light-likeness condition, something special happens. Indeed, consider the explicit form \(\xi_{i j}\) with \(i = 1, \dots, 4\), \(j = 1,2\). Then,
\begin{equation}
  \xi \begin{pmatrix}
    0 & -c_2\\
    c_1 & 0
  \end{pmatrix} + c
  \begin{pmatrix}
    \epsilon & 0\\0 & \epsilon
  \end{pmatrix}
  \xi^* =
  \begin{pmatrix}
    c_1 \xi_{12} + c \xi_{21}^* & -c_2 \xi_{11} + c \xi_{22}^*\\
    c_1 \xi_{22} - c \xi_{11}^* & -c_2 \xi_{21} - c \xi_{12}^*\\
    c_1 \xi_{32} + c \xi_{41}^* & -c_2 \xi_{31} + c \xi_{42}^*\\
    c_1 \xi_{42} - c \xi_{31}^* & -c_2 \xi_{41} - c \xi_{32}^*
  \end{pmatrix}.
\end{equation}
If we set
\begin{gather}
  \xi_{11} = \frac {c_1}{c} \xi_{22}^* = \frac {c}{c_2} \xi_{22}^*,\\
  \xi_{21} = -\frac {c_1}{c} \xi_{12}^* = -\frac {c}{c_2} \xi_{12}^*,\\
  \xi_{31} = \frac {c_1}{c} \xi_{42}^* = \frac {c}{c_2} \xi_{42}^*,\\
  \xi_{41} = -\frac {c_1}{c} \xi_{32}^* = -\frac {c}{c_2} \xi_{32}^*,
\end{gather}
then the left-hand side vanishes.  If we want to see the transformation
\begin{equation}
  \xi \mapsto   \xi \begin{pmatrix}
    0 & -c_2\\
    c_1 & 0
  \end{pmatrix} + c
  \begin{pmatrix}
    \epsilon & 0\\
    0 & \epsilon
  \end{pmatrix} \xi^*
\end{equation}
as a real-linear operator acting on a \(16\)-dimensional real space (the space of \(\xi\)), then this linear operator has a real \(8\)-dimensional kernel.  Therefore, it must have a real \(8\)-dimensional image.

If we fix the normalization of \(\xi\) by \(\vert \xi_{12}\vert^2 + \vert \xi_{22}\vert^2 + \vert \xi_{32}\vert^2 + \vert \xi_{42}\vert^2 = c_2\) and as a consequence \(\vert \xi_{11}\vert^2 + \vert \xi_{21}\vert^2 + \vert \xi_{31}\vert^2 + \vert \xi_{41}\vert^2 = c_1\), then we have
\begin{equation}
  \xi^\dagger \xi =
  \begin{pmatrix}
    c_1 & 0\\0 & c_2
  \end{pmatrix}
\end{equation}
and therefore
\begin{equation}
  \lambda^\dagger \lambda = p,
\end{equation}
where we recall that \(\lambda\) is a \(2 \times 4\) matrix.  The hermiticity of \(p\) is manifest.

Next we define
\begin{equation}
  \tilde{\xi} =
  \begin{pmatrix}
    c_2 \alpha \xi_{11} & c_1 \alpha \xi_{12}\\
    c_2 \alpha \xi_{21} & c_1 \alpha \xi_{22}\\
    c_2 \beta \xi_{31} & c_1 \beta \xi_{32}\\
    c_2 \beta \xi_{41} & c_1 \beta \xi_{42}
  \end{pmatrix},
\end{equation}
with
\begin{gather}
  \alpha = \frac 1 c \sqrt{\frac {\vert\xi_{32}\vert^2 + \vert\xi_{42}\vert^2}{\vert\xi_{12}\vert^2 + \vert\xi_{22}\vert^2}}, \qquad
  \beta = -\frac 1 c \sqrt{\frac {\vert\xi_{12}\vert^2 + \vert\xi_{22}\vert^2}{\vert\xi_{32}\vert^2 + \vert\xi_{42}\vert^2}}.
\end{gather}
The \(\xi\) and \(\tilde{\xi}\) satisfy some useful identities
\begin{gather}
  \xi^\dagger \tilde{\xi} = 0, \qquad
  \ast (\xi^* \epsilon \xi^\dagger) = \tilde{\xi} \epsilon \tilde{\xi}^\trans,
\end{gather}
where \(\ast M\) computes the Hodge dual of a \(4 \times 4\) anti-symmetric matrix \(M\), i.e.\ \((\ast M)_{a b} = \frac 1 2 \epsilon_{a b c d} M^{c d}\).

We can also easily compute that
\begin{gather}
  \tilde{\xi}^\dagger \tilde{\xi} =
  \begin{pmatrix}
    c_2 & 0\\0 & c_1
  \end{pmatrix}, \qquad
  \xi \epsilon \xi^\trans + \tilde{\xi} \epsilon \tilde{\xi}^\trans = c
\begin{pmatrix}
  \epsilon & 0\\
  0 & \epsilon
\end{pmatrix}.
\end{gather}

Using the definitions
\begin{gather}
  \lambda = U \xi v^\dagger, \qquad
  \tilde{\lambda} = U \tilde{\xi} v^\dagger,
\end{gather}
and the identities
\begin{equation}
  v^\trans \epsilon v = \epsilon, \qquad
  \ast (U^* M U^\dagger) = U (\ast M) U^\trans,
\end{equation}
valid for \(v \in \grp{SU}(2)\), \(U \in \grp{SU}(4)\) and \(M\) anti-symmetric, we can show that
\begin{gather}
  \lambda^\dagger \tilde{\lambda} = 0, \qquad
  \ast (\lambda^* \epsilon \lambda^\dagger) = \tilde{\lambda} \epsilon \tilde{\lambda}^\trans,\\
  p = \lambda^\dagger \lambda, \qquad \bar{p} = -\epsilon \lambda^\trans \lambda^* \epsilon,\\
  q = \lambda \epsilon \lambda^\trans + \tilde{\lambda} \epsilon \tilde{\lambda}^\trans, \qquad
  \bar{q} = \lambda^* \epsilon \lambda^\dagger + \tilde{\lambda}^* \epsilon \tilde{\lambda}^\dagger.
\end{gather}
The constraints in the first line above are precisely the pure spinor constraints in a 4D-reduced form.  They were presented in this form by Berkovits and Fleury in ref.~\cite{Berkovits_2013}.

We can now do the construction in the opposite direction; we start with two \(2 \times 4\) matrices \(\lambda\) and \(\tilde{\lambda}\) satisfying the pure spinor constraints, we compute \(p\), \(\bar{p}\), \(q\) and \(\bar{q}\) by the equations above and show that they satisfy the correct constraints.  A similar construction was done by Howe in ref.~\cite{Howe1991}.

Clearly \(q\) and \(\bar{q}\) are anti-symmetric and also \(\bar{q} = q^*\) and \(\bar{q} = \ast q\).  To check that the other constraints are satisfied by \(q\) and \(\bar{q}\) we need the following identity
\begin{equation}
  \label{eq:hodge-id}
  M N + (\ast N)(\ast M) = \frac 1 2 \tr(M N)\; \mathbf{1}_4,
\end{equation}
where \(M\), \(N\) are \(4 \times 4\) anti-symmetric matrices.  This identity can be shown by explicit computation.

There are four terms in the product \(q \bar{q}\):
\begin{equation}
  q \bar{q} = (\lambda \epsilon \lambda^\trans)(\lambda^* \epsilon \lambda^\dagger) +
  \ast (\lambda^* \epsilon \lambda^\dagger) \ast (\lambda \epsilon \lambda^\trans) +
  (\lambda \epsilon \lambda^\trans) \ast(\lambda \epsilon \lambda^\trans) +
  (\tilde{\lambda} \epsilon \tilde{\lambda}^\trans) \ast(\tilde{\lambda} \epsilon \tilde{\lambda}^\trans).
\end{equation}
The last two terms vanish by anti-symmetry while the first two terms combine in a trace by using eq.~\eqref{eq:hodge-id}.  We obtain
\begin{equation}
  q \bar{q} = \frac 1 2 \tr((\lambda \epsilon \lambda^\trans)(\lambda^* \epsilon \lambda^\dagger)) \; \mathbf{1}_4 = - \det(\lambda^\dagger \lambda) \; \mathbf{1}_4 = -\det p \; \mathbf{1}_4.
\end{equation}

The constraints for \(p\) are automatically satisfied.

\section{Supersymmetrization}
\label{sec:supersymmetrization}

The infinitesimal supersymmetry generator is represented by the super-matrix
\begin{equation}
  m =
  \begin{pmatrix}
    0 & \theta & 0\\
    0 & 0 & \bar{\theta}\\
    0 & 0 & 0
  \end{pmatrix},
\end{equation}
where we have adopted a \((2\vert 4\vert 2) \times (2\vert 4\vert 2)\) block writing.  The corresponding finite supersymmetry transformation is
\begin{equation}
  M = \exp(m) =
  \begin{pmatrix}
    1 & \theta & \tfrac 1 2 \theta \bar{\theta}\\
    0 & 1 & \bar{\theta}\\
    0 & 0 & 1
  \end{pmatrix}.
\end{equation}

In fact, we can combine the supersymmetry transformations with translations in a single super-matrix
\begin{equation}
  M =
  \begin{pmatrix}
    1 & \theta & x^-\\
    0 & 1 & \bar{\theta}\\
    0 & 0 & 1
  \end{pmatrix}, \qquad
  M^{-1} =
  \begin{pmatrix}
    1 & -\theta & -x^+\\
    0 & 1 & -\bar{\theta}\\
    0 & 0 & 1
  \end{pmatrix},
\end{equation}
where \(x^\pm = x \mp \frac 1 2 \theta \bar{\theta}\).

The supersymmetrized versions of the bosonic quantities, can be obtained by acting with the transformation matrix \(M\).  For example, we find
\begin{equation}
  M
  \begin{pmatrix}
    0\\0\\1
  \end{pmatrix} =
  \begin{pmatrix}
    x^-\\ \bar{\theta}\\ 1
  \end{pmatrix}, \qquad
  \begin{pmatrix}
    0 & 0 & 1
  \end{pmatrix} M^{-1} =
  \begin{pmatrix}
    1 & -\theta & -x^+
  \end{pmatrix}.
\end{equation}
These quantities are supersymmetric but not superconformal.  Under superconformal action they transform as elements of a coset, i.e.\ we need to include a right action by the stability group in order to preserve their form.

We want to supersymmetrize the bosonic variables discussed in the previous section.  Using their transformation properties under Lorentz and \(R\)-symmetry, it is easy to identify the bosonic part which should be supersymmetrized.  We find
\begin{gather}
  \begin{pmatrix}
    0 & 0\\
    \lambda & 0\\
    0 & 1
  \end{pmatrix} \to M \begin{pmatrix}
    0 & 0\\
    \lambda & 0\\
    0 & 1
  \end{pmatrix} =
  \mathcal{U} = \begin{pmatrix}
    \theta \lambda & x^-\\
    \lambda & \bar{\theta}\\
    0 & 1
  \end{pmatrix},\\
  \begin{pmatrix}
    0 & \lambda^\dagger & 0\\
    1 & 0 & 0\\
  \end{pmatrix} \to \begin{pmatrix}
    0 & \lambda^\dagger & 0\\
    1 & 0 & 0\\
  \end{pmatrix} M^{-1} =
  \mathcal{V} = \begin{pmatrix}
    0 & \lambda^\dagger & -\lambda^\dagger \bar{\theta}\\
    1 & -\theta & -x^+
  \end{pmatrix}.
\end{gather}
This defines \(\mathcal{U}\) to be a \((2\vert 4\vert 2) \times (2\vert 2)\) super-matrix and \(\mathcal{V}\) to be a \((2 \times 2) \times (2 \vert 4\vert 2)\) super-matrix.  These quantities transform covariantly under supersymmetry transformations (\(\mathcal{U}\) under a left action and \(\mathcal{V}\) under a right action), but under general superconformal transformations they require a compensating transformation by a \((2 \vert 2) \times (2 \vert 2)\) super-matrix (on the right for \(\mathcal{U}\) and on the left for \(\mathcal{V}\)).

The spinor \(\tilde{\lambda}\) has the same transformation properties as \(\lambda\) so we can define \(\tilde{\mathcal{U}}\) and \(\tilde{\mathcal{V}}\) by replacing \(\lambda \to \tilde{\lambda}\) in the definition of \(\mathcal{U}\) and \(\mathcal{V}\), respectively.

Here are the possible \(\mathcal{V} \mathcal{U}\) products which yield \((2 \vert 2) \times (2 \vert 2)\) super-matrices
\begin{gather}
  \mathcal{V} \tilde{\mathcal{U}} = 0, \qquad \tilde{\mathcal{V}} \mathcal{U} = 0,\\
  \mathcal{V} \mathcal{U} =
  \begin{pmatrix}
    p & 0\\0 & 0
  \end{pmatrix}, \qquad
  \tilde{\mathcal{V}} \tilde{\mathcal{U}} =
  \begin{pmatrix}
    \tilde{\lambda}^\dagger \tilde{\lambda} & 0\\0 & 0
  \end{pmatrix}.
\end{gather}
The identities in the first line above are consequences of the constraints \(\lambda^\dagger \tilde{\lambda} = 0\) and \(\tilde{\lambda}^\dagger \lambda = 0\) described previously.  We will describe later how the remaining pure spinor constraints arise.

The product \(\mathcal{V}_i \mathcal{U}_j\) between such super-matrices corresponding to different points in superspace is
\begin{equation}
  \mathcal{V}_i \mathcal{U}_j =
  \begin{pmatrix}
    \lambda_i^\dagger \lambda_j & -\lambda_i^\dagger \bar{\theta}_{i j}\\
    -\theta_{i j} \lambda_j & -x_{i j}^{+-}
  \end{pmatrix},
\end{equation}
where \(\theta_{i j} = \theta_i - \theta_j\), \(x_{i j}^{+-} = x_i^+ - x_j^- - \theta_i \bar{\theta}_j\).

Under superconformal transformations all these quantities transform homogeneously, and get multiplied by a left and a right super-matrix. This implies that the super-determinant is covariant under superconformal transformations.

%%%%%%%%%%%%%%%%%%%%%%%%%%%%%%%%%%%%%%%%%%%%%%%%%%%%%%%%%%%%%%%%%%%%%%
\section{Correlation functions}
\label{sec:corr-funct}

Using the expression for \(q\) and \(\bar{q}\) in terms of pure spinor components, we find
\begin{equation}
  \tr(q_1 \bar{q}_2) = -2 (\det (\lambda_2^\dagger \lambda_1) + \det (\tilde{\lambda}_2^\dagger \lambda_1) + \det (\lambda_2^\dagger \tilde{\lambda}_1) + \det (\tilde{\lambda}_2^\dagger \tilde{\lambda}_1)).
\end{equation}

Supersymmetrically we expect the answer for the correlation function of two scalar superfields \(\langle q_1 \cdot \Phi(1)\; \bar{q}_2 \cdot \bar{\Phi}(2)\rangle\) to be obtained by replacing
\begin{equation}
   \frac{\det(\lambda_2^\dagger \lambda_1)}{\det x_{12}} \to \sdet
  \begin{pmatrix}
    \lambda_2^\dagger \lambda_1 & -\lambda_2^\dagger \bar{\theta}_{21}\\
    -\theta_{21} \lambda_1 & -x_{21}^{+-}
  \end{pmatrix} = \sdet (\mathcal{V}_2 \mathcal{U}_1),
\end{equation}
where
\begin{equation}
  \sdet
  \begin{pmatrix}
    M & N\\P & Q
  \end{pmatrix} = \frac {\det (M - N Q^{-1} P)}{\det Q} = \frac {\det M}{\det (Q - P M^{-1} N)}.
\end{equation}
This is similar to the result obtained in ref.~\cite{Howe1997a}, but it differs in some details.  In the next section we will discuss the relation to harmonic superspace in more detail.

Explicitly, we find
\begin{multline}
  \sdet (\mathcal{V}_2 \mathcal{U}_1) = \frac {\det (\lambda_2^\dagger \lambda_1 + \lambda_2^\dagger \bar{\theta}_{21} x_{21}^{+-,-1} \theta_{21} \lambda_1)}{\det x_{21}^{+-}} =\\
  =-\frac 1 2 \frac {\tr (\lambda_2^\dagger (1 + \bar{\theta}_{21} x_{21}^{+-,-1} \theta_{21}) \lambda_1 \epsilon \lambda_1^\trans (1 + \bar{\theta}_{21} x_{21}^{+-,-1} \theta_{21})^\trans \lambda_2^* \epsilon)}{\det x_{21}^{+-}},
\end{multline}
where we have used the identity \(\det x = -\tfrac 1 2 \tr(x \epsilon x^\trans \epsilon)\), valid for all \(2 \times 2\) matrices.

Adding up all the four contributions reproduces \(q_1 = \lambda_1 \epsilon \lambda_1^\trans + \tilde{\lambda}_1 \epsilon \tilde{\lambda}_1^\trans\) and \(\bar{q}_2 = \lambda_2^* \epsilon \lambda_2^\dagger + \tilde{\lambda}_2^* \epsilon \tilde{\lambda}_2^\dagger\).

Other superconformal covariants can be formed by joining together two \((2 \vert 4\vert 2) \times (2 \vert 2)\) super-matrices to form a \((2 \vert 4\vert 2) \times (2 \vert 4\vert 2)\) super-matrix.  We define
\begin{equation}
  \sdet (\mathcal{U}_1; \mathcal{U}_2) = \sdet
  \begin{pmatrix}
    x_1^- & \theta_1 \lambda_1 & \theta_2 \lambda_2 & x_2^-\\
    \bar{\theta}_1 & \lambda_1 & \lambda_2 & \bar{\theta}_2\\
    1 & 0 & 0 & 1
  \end{pmatrix},
\end{equation}
where we have permuted the columns in order to make it possible to compute the super-determinant.\footnote{In order to be able to compute the super-determinant the super-matrix should be a linear map between spaces of the same structure, i.e.\ the sequence of integers describing the dimensions of the even and odd spaces should be the same for the domain of definition and for the image.}  In general permuting rows or columns of different parity does not preserve the value of the super-determinant, but in the expression above we will simply take the right-hand side to be definition of the \(\sdet (\mathcal{U}_1; \mathcal{U}_2)\) notation.

The easiest way to compute this super-determinant is to apply a super-translation which preserves it
\begin{equation}
  \sdet
  \begin{pmatrix}
    x_1^- & \theta_1 \lambda_1 & \theta_2 \lambda_2 & x_2^-\\
    \bar{\theta}_1 & \lambda_1 & \lambda_2 & \bar{\theta}_2\\
    1 & 0 & 0 & 1
  \end{pmatrix} =
  \sdet
  \begin{pmatrix}
    0 & 0 & -\theta_{12} \lambda_2 & -x_{12}^{+-}\\
    0 & \lambda_1 & \lambda_2 & -\bar{\theta}_{12}\\
    1 & 0 & 0 & 1
  \end{pmatrix}.
\end{equation}

Next, we apply the identity
\begin{equation}
  \sdet
  \begin{pmatrix}
    M & N\\P & Q
  \end{pmatrix} = \sdet (M - N Q^{-1} P) \sdet Q^{-1},
\end{equation}
where \(M\), \(N\), \(P\) and \(Q\) are themselves super-matrices (see ref.\cite[page 104, theorem 3.4]{MR914369} for a proof).  This yields
\begin{multline}
  \sdet (\mathcal{U}_1; \mathcal{U}_2) = \sdet
  \left(\begin{pmatrix}
    0 & 0 & -\theta_{12} \lambda_2\\
    0 & \lambda_1 & \lambda_2
  \end{pmatrix} -
  \begin{pmatrix}
    1 & 0 & 0
  \end{pmatrix}
  \begin{pmatrix}
    -x_{12}^{+-}\\ -\bar{\theta}_{12}
  \end{pmatrix}\right) =\\= \frac {\det x_{12}^{+-}}{\det (\lambda_1, (1 + \bar{\theta}_{12} x_{12}^{+-,-1} \theta_{12}) \lambda_2)}.
\end{multline}

Finally, we need to compute the \(4 \times 4\) determinant in the denominator.  For a \(4 \times 4\) matrix \(M = (\lambda, \mu)\) where \(\lambda\), \(\mu\) are \(4 \times 2\) matrices, we have
\begin{multline}
  \det M = \epsilon_{a b c d} M^a{}_1 M^b{}_2 M^c{}_3 M^d{}_4 =
  \epsilon_{a b c d} \lambda^a{}_1 \lambda^b{}_2 \mu^c{}_1 \mu^d{}_2 =\\=
  \frac 1 4 \epsilon_{a b c d} (\lambda \epsilon \lambda^\trans)^{a b} (\mu \epsilon \mu^\trans)^{c d} =
  -\frac 1 2 \tr(\ast (\lambda \epsilon \lambda^\trans) (\mu \epsilon \mu^\trans)).
\end{multline}
Using this formula and applying the constraint \(\ast (\lambda \epsilon \lambda^\trans) = \tilde{\lambda}^* \epsilon \tilde{\lambda}^\dagger\), we find
\begin{equation}
  \sdet (\tilde{\mathcal{V}}_1 \mathcal{U}_2) = \sdet (\mathcal{U}_1; \mathcal{U}_2)^{-1}.
\end{equation}
This identity is the supersymmetric version of the second pure spinor constraint.  Since this holds for arbitrary \(\mathcal{U}_2\), the dependence on \(\mathcal{U}_2\) can be eliminated and then the identity reads
\begin{equation}
  \sdet (\tilde{\mathcal{V}}\; \cdot) = \sdet (\mathcal{U}; \cdot)^{-1},
\end{equation}
where \(\cdot\) can be replaced by any \((2\vert 4\vert 2) \times (2 \vert 2)\) super-matrix.

Let us check that the expression for the correlation function
\begin{equation}
  \langle q_1 \cdot \Phi(1)\; \bar{q}_2 \cdot \bar{\Phi}(2)\rangle \propto \sdet (\mathcal{V}_1 \mathcal{U}_2) + \sdet (\mathcal{V}_1 \tilde{\mathcal{U}}_2) + \sdet (\tilde{\mathcal{V}}_1 \mathcal{U}_2) + \sdet (\tilde{\mathcal{V}}_1 \tilde{\mathcal{U}}_2)
\end{equation}
has the required symmetry properties.  For example, since \(q \cdot \Phi = \bar{q} \cdot \bar{\Phi}\), we should find that \(\langle q_1 \cdot \Phi(1)\; \bar{q}_2 \cdot \bar{\Phi}(2)\rangle = \langle \bar{q}_1 \cdot \bar{\Phi}(1)\; q_2 \cdot \Phi(2)\rangle\).  In sec.~\ref{sec:harmonic-superspace} we show that \(\sdet (\mathcal{V}_i \tilde{\mathcal{U}}_j) = \sdet X_{i j}\) with \(X_{i j} = X_i - X_j\) and therefore
\begin{equation}
  \sdet (\mathcal{V}_i \tilde{\mathcal{U}}_j) = \sdet X_{i j} = \sdet (-X_{j i}) = \sdet (-\mathcal{V}_j \tilde{\mathcal{U}}_i) = \sdet (\mathcal{V}_j \tilde{\mathcal{U}}_i).
\end{equation}
Moreover,
\begin{equation}
  \sdet (\mathcal{V}_i \mathcal{U}_j) = \sdet (\tilde{\mathcal{U}}_i; \mathcal{U}_j)^{-1} = \sdet (\mathcal{U}_j; \tilde{\mathcal{U}}_i)^{-1} = \sdet (\tilde{\mathcal{V}}_j \tilde{\mathcal{U}}_i).
\end{equation}
The required symmetry follows from these two identities.

\section{Harmonic superspace}
\label{sec:harmonic-superspace}

In this section we drop the reality conditions and complexify all coordinates.

The results in the previous section can also be presented in the language of harmonic superspace, but we need to use \emph{two} copies. The relevant superspace is the \(G_{2\vert 2}(4\vert 4)\) super-Grassmannian introduced by Howe and Hartwell in ref.~\cite{Howe1995}.  The super-Grassmannian \(G_{2\vert 1}(4\vert 2)\) was introduced even earlier for \(\mathcal{N}=2\) superconformal field theories by Rosly and Schwarz (see ref.~\cite{Rosly1986}). See also the book~\cite{MR1632008} by Manin, for a more mathematical treatment.

For a \(4 \times 4\) anti-symmetric matrix, the action of the Hodge dual can be written as
\begin{equation}
  \ast
  \begin{pmatrix}
    M & N\\-N^\trans & P
  \end{pmatrix} =
  \begin{pmatrix}
    P & \epsilon N \epsilon\\
    -\epsilon N^\trans \epsilon & M
  \end{pmatrix}.
\end{equation}

Next assume that we can gauge-fix
\begin{equation}
  \lambda =
  \begin{pmatrix}
    1 \\ y^\dagger
  \end{pmatrix}, \qquad
  \lambda^\dagger =
  \begin{pmatrix}
    1 & y
  \end{pmatrix}.
\end{equation}
Then,
\begin{equation}
  \lambda \epsilon \lambda^\trans =
  \begin{pmatrix}
    \epsilon & \epsilon y^*\\
    y^\dagger \epsilon & y^\dagger \epsilon y^*
  \end{pmatrix}, \qquad
  \ast (\lambda^* \epsilon \lambda^\dagger) =
  \begin{pmatrix}
    y^\trans \epsilon y & -y \epsilon\\
    -\epsilon y^\trans & \epsilon
  \end{pmatrix} =
  \tilde{\lambda} \epsilon \tilde{\lambda}^\trans,
\end{equation}
where
\begin{equation}
  \tilde{\lambda} =
  \begin{pmatrix}
    -y\\1
  \end{pmatrix}, \qquad
  \tilde{\lambda}^\dagger =
  \begin{pmatrix}
    -y^\dagger & 1
  \end{pmatrix}.
\end{equation}
The equality \(\ast (\lambda^* \epsilon \lambda^\dagger) = \tilde{\lambda} \epsilon \tilde{\lambda}^\trans\) relies on the identity \(y \epsilon y^\trans = y^\trans \epsilon y\), valid for \(2 \times 2\) matrices.  This implies
\begin{gather}
  q =
  \begin{pmatrix}
    \epsilon + y^\trans \epsilon y & \epsilon y^* - y \epsilon\\
    y^\dagger \epsilon - \epsilon y^\trans & y^\dagger \epsilon y^* + \epsilon
  \end{pmatrix}, \qquad
  \bar{q} =
  \begin{pmatrix}
    \epsilon + y^\dagger \epsilon y^* & \epsilon y - y^* \epsilon\\
    y^\trans \epsilon - \epsilon y^\dagger & y^\trans \epsilon y + \epsilon
  \end{pmatrix}, \qquad
  p = 1 + y y^\dagger.
\end{gather}

Using this gauge-fixed form of the variables \(\lambda\) and \(\lambda^\dagger\) we find
\begin{equation}
  \mathcal{V} =
  \begin{pmatrix}
    0 & 1 & y & -\bar{\theta}' - y \bar{\theta}''\\
    1 & -\theta' & -\theta'' & -x^+
  \end{pmatrix},
\end{equation}
where we split the \(\theta\) and \(\bar{\theta}\) in \(2 \times 2\) blocks \(\theta = (\theta', \theta'')\), \(\bar{\theta} = \left(\begin{smallmatrix}\bar{\theta}'\\ \bar{\theta}''\end{smallmatrix}\right)\).  By an \(\grp{SL}(2\vert 2)\) left action the \(\mathcal{V}\) super-matrix can be transformed into a canonical form
\begin{equation}
  \mathcal{V} \sim \begin{pmatrix}
    0 & 1 & y & -\bar{\theta}' - y \bar{\theta}''\\
    1 & 0 & \theta' y - \theta'' & -\theta' \bar{\theta}' - \theta' y \bar{\theta}'' - x^+
  \end{pmatrix} \equiv \left(\begin{matrix}
      0 & 1\\
      1 & 0
\end{matrix}\mathrel{\bigg|} X\right),
\end{equation}
where
\begin{equation}
  X =
  \begin{pmatrix}
    y & -\bar{\theta}' - y \bar{\theta}''\\
    \theta' y - \theta'' & -x -\tfrac 1 2 \theta' \bar{\theta}' + \tfrac 1 2 \theta'' \bar{\theta}'' - \theta' y \bar{\theta}''
  \end{pmatrix}.
\end{equation}

A similar analysis can be done for the \(\tilde{\mathcal{U}}\) and its canonical form contains the same components as \(X\).  We have
\begin{equation}
  \mathcal{V}_i \tilde{\mathcal{U}}_j = X_i - X_j = X_{i j}.
\end{equation}
It is important to notice that the product in the left-hand side can be written as a difference.  These differences are manifestly invariant under a \(\grp{GL}(2\vert 2)\) subgroup of the superconformal group, which acts by translations on \(X\).  This is one advantage of this parametrization of the harmonic variables.

In ref.~\cite{Howe1997a}, Howe and West expressed the correlation function of \emph{projections} of scalar superfields as a super-determinant of \(X_{i j}\).  More explicitly, instead of being contracted with the vector \(q\), in their work the superfield \(\Phi\) was contracted with \(\lambda \epsilon \lambda^\trans\), so the second term \(\tilde{\lambda} \epsilon \tilde{\lambda}^\trans\) did not appear.  For our purposes, it is essential that we keep all the contributions since the super-Wilson loop contains contractions \(q \cdot \Phi\) for generic \(q\) while in ref.~\cite{Howe1997a} the superfield \(\Phi\) is always contracted with a rank two matrix.

In order to add these contributions, we follow the same idea as above and, after putting \(\tilde{\mathcal{V}}\) and \(\mathcal{U}\) in canonical forms, we are led to introduce \(\tilde{X}\) as
\begin{equation}
  \tilde{X} =
  \begin{pmatrix}
    -y^\dagger & y^\dagger \bar{\theta}' - \bar{\theta}''\\
    -\theta' - \theta'' y^\dagger & -x + \tfrac 1 2 \theta' \bar{\theta}' - \frac 1 2 \theta'' \bar{\theta}'' + \theta'' y^\dagger \bar{\theta}'
  \end{pmatrix}.
\end{equation}
These are coordinates on a second \(G_{2\vert 2}(4\vert 4)\) super-Grassmannian.

So far we have shown that two of the terms in the scalar--scalar superfield correlation function can be written in a simple way in terms of harmonic coordinates
\begin{gather}
  \sdet \mathcal{V}_1 \tilde{\mathcal{U}}_2 = \sdet X_{12}, \qquad
  \sdet \tilde{\mathcal{V}}_1 \mathcal{U}_2 = \sdet \tilde{X}_{12}.
\end{gather}
The other two contributions \(\sdet \mathcal{V}_1 \mathcal{U}_2\) and \(\sdet \tilde{\mathcal{V}}_1 \tilde{\mathcal{U}}_2\) do \emph{not} have such a simple form.  In particular, they can not be written in terms of differences of coordinates.  To see this it's easiest to notice that \(\mathcal{V} \mathcal{U} \neq 0\).

The \(\grp{GL}(2\vert 2)\) symmetry group which acts by translating the variables \(X\) has appeared before in ref.~\cite[sec.~3.1]{Berkovits2008a} by Berkovits and Maldacena on fermionic T-duality.  The main result of this paper is that the type IIB \(\text{AdS}_5 \times S^5\) string background is preserved by a sequence of T-dualities in the four space-time directions and in the eight fermionic directions corresponding to a set of anti-commuting chiral supercharges.  Moreover, this invariance under T-duality implies Yangian symmetry of the super-Wilson loop (see also ref.~\cite{Beisert:2008iq}).

As shown in ref.~\cite{Beisert:2015uda} the non-chiral super-Wilson loop with a coupling to scalars is also Yangian invariant to one loop order.  Is it possible to show Yangian invariance to all orders in the 't Hooft coupling by applying T-duality as in the chiral case? Berkovits and Maldacena also point out in the sections~3.1 and 3.2 of ref.~\cite{Berkovits2008a} that there is \emph{another} set of T-dualities which also preserve the \(\text{AdS}_5 \times S^5\) background.  The directions in which these T-dualities are performed are parametrized by \(X\) but can also correspond to \(\tilde{X}\). We should emphasize that \(X\) and \(\tilde{X}\) do not transform in the same way and in fact translations of \(X\) and of \(\tilde{X}\) do not commute.

An obvious guess is that one needs to perform the T-dualities corresponding to \(X\) and then the T-dualities corresponding to \(\tilde{X}\), or the other way around.  At weak coupling and for the chiral case, the effect of T-dualities was to pass to the dual space \(p_i \to x_i - x_{i+1}\), \(q_i \to \theta_i - \theta_{i+1}\) and one may hope to introduce a dual space for the non-chiral superspace in a similar way.  The strong-coupling situation is more confusing.  For example, in the chiral case one starts with some D3-branes which after four T-dualities along the space-time directions become D(-1) instantons, which are the vertices of the super-Wilson loop.  In the non-chiral case we can perform some T-dualities along four space-time directions in \(X\) and then four T-dualities along the four space-time directions in \(\tilde{X}\).  These space-time directions, however, differ by nilpotent quantities.  It should be interesting to understand what kind of brane configuration gets generated by such a sequence of T-dualities.

\bibliographystyle{plain}
\bibliography{ps}

\end{document}